\documentclass{aa}
\usepackage{txfonts}
\usepackage{rotating}
\usepackage{graphicx}
\usepackage{natbib}
\usepackage{textcomp}

\bibpunct{(}{)}{;}{a}{}{,}

\usepackage{color}

\newcommand{\Msun}{M_\odot}

\newcommand{\as}{\ifmmode {^{\scriptscriptstyle\prime\prime}}
        \else $^{\scriptscriptstyle\prime\prime}$\fi}

\begin{document}
\title{Chemistry in Disks VIII:\\
the CS molecule as an analytic tracer of turbulence in disks.
\thanks{Based on observations carried out with the IRAM Plateau de Bure interferometer.
 IRAM is supported by INSU/CNRS (France), MPG (Germany) and IGN (Spain).}
}

\author{St\'ephane Guilloteau \inst{1,2}, Anne Dutrey\inst{1,2}, Valentine Wakelam\inst{1,2},  Franck Hersant\inst{1,2},
Dmitry Semenov \inst{3}, Edwige Chapillon\inst{4}, Thomas Henning \inst{3},  Vincent Pi\'etu \inst{5}}
%
\institute{
Univ. Bordeaux, LAB, UMR 5804, F-33270, Floirac, France
\and
CNRS, LAB, UMR 5804, F-33270 Floirac, France\\
  \email{[guilloteau,dutrey,wakelam,hersant]@obs.u-bordeaux1.fr}
\and
Max-Planck-Institut f\"ur Astronomie, K\"onigstuhl 17, D-69117
Heidelberg, Germany
\and
Academia Sinica Institute of Astronomy and Astrophysics, P.O. Box 23-141, Taipei 106, Taiwan, R.O.C.
\and
IRAM, 300 rue de la piscine, F-38406
Saint Martin d'H\`eres, France
}

\offprints{S.Guilloteau, \email{Stephane.Guilloteau@obs.u-bordeaux1.fr}}

\date{Received 4 Sep 2012 / Accepted 14 Oct 2012} %
\authorrunning{Guilloteau et al.} %
\titlerunning{Turbulence in disks: the CS tool.}

\abstract
{Turbulence is thought to be a key driver of the evolution of protoplanetary disks, regulating the mass accretion process,
the transport of angular momentum, and the growth of dust particles.}
{We intend to determine the magnitude of the turbulent motions in the outer parts ($> 100$ AU) of the disk surrounding DM Tau.}
{Turbulent motions can be constrained by measuring the nonthermal broadening of line emission
from heavy molecules. We used the IRAM Plateau de Bure interferometer to study emission from the CS molecule
in the disk of DM Tau. High spatial
($1.4 \times 1 \farcs$) and spectral resolution (0.126~km\,s$^{-1}$) CS J=3-2 images
provide constraints on the molecule distribution and velocity structure of the disk.
A low sensitivity CS J=5-4 image was used in conjunction to evaluate the excitation conditions.
We analyzed the data in terms of two parametric disk models, and compared the results with
detailed time-dependent chemical simulations.}
{The CS data confirm the relatively low temperature suggested by observations of other simple molecules.
The intrinsic linewidth derived from the CS J=3-2 data is much larger than expected from pure thermal broadening.
The magnitude of the derived nonthermal component depends only weakly on assumptions about the location of the CS molecules
with respect to the disk plane. Our results indicate turbulence with a Mach number around 0.4 -- 0.5 in the molecular
layer. Geometrical constraints suggest that this layer is located near one scale height, in reasonable agreement
with chemical model predictions.}
{}


\maketitle{}

\section{Introduction}

Turbulence is thought to be a major actor throughout the lifetime of
protoplanetary disks, by regulating  the mass accretion and angular
momentum transport, planetesimal formation and planet migration,
and mixing processes relevant for gas and grain chemistry
\citep[see, e.g.][]{Henning_2008}.
The magneto-rotational instability \citep[MRI,][]{Balbus+Hawley_1991} is currently the most widely accepted
theory for an origin of the effective turbulent viscosity. The
ionization fraction in the outer disk is sufficient to allow
for MRI to operate, but there may be a ``dead'' laminar zone in the inner ($<$ 10 AU) disk
interior,  efficiently shielded from ionizing
radiation \citep{Gammie_1996}. Its extent depends on a
variety of factors including magnetic field morphology, X-ray activity, abundance of metals,
and dust properties \citep{Ilgner+Nelson_2006}.
The 2D disk models by \citet{Keller+Gail_2004} and \citet{Tscharnuter+Gail_2007}
 show the possible presence of large-scale circulation
\citep{Urpin_1984, Regev+Gitelman_2002}, but such motions would
be suppressed by turbulence on the scale of the disk thickness.
These meridional flows may be artifacts of the 2D-hydrodynamical disk models with fixed $\alpha$ viscosity.
More realistic 3D MHD simulations do not show preferential directions of the large-scale gas-flows
\citep[see e.g.,][]{Flock+etal_2011} 

%
Turbulence may also provide a source of relative velocities
for grain coagulation to occur \citep[e.g.][]{Voelk+etal_1980,Beckwith+etal_2000}.
The interplay between gas and solid particle motion,
would also lead to high local ``dust'' overdensities and
planetesimal formation by gravitational collapse \citep{Johansen+etal_2007,
Johansen+Klahr+Henning_2011}.
The turbulent state may have a strong
impact on the migration efficiency \citep{Papaloizou+Terquem_2006,Oishi+etal_2007,Uribe+etal_2011}.
Finally, models of gas-phase chemistry including surface chemistry on
grains are sensitive to the level of turbulent mixing \citep[e.g.][]{
Ilgner+etal_2004, Ilgner+Nelson_2006,Semenov+Wiebe_2011} which may partially explain the presence
of cold CO in disks \citep{Dartois+etal_2003, Semenov+etal_2006, Aikawa_2007, Hersant+etal_2009}.


Despite this enormous importance of turbulence for disks, even its
magnitude is poorly constrained. Most constraints on the
anomalous viscosity parameter $\alpha$ come from measured accretion rates
onto the star, which is mostly determined by inner disk properties.
\citet{Hueso+Guillot_2005}
have attempted to constrain
the overall value of the $\alpha$ parameter by simulations of the
viscous evolution of disks as a function of time. More recent results
on viscous spreading of dust disks from \citet{Guilloteau+etal_2011}
suggest
$\alpha$ decreases with time, with values less than $0.001$ near 100 AU
after a few Myr.

Direct measurements of the nonthermal gas motions resulting from mesoscopic
scale turbulence remain rare. \citet{Carr+etal_2004} report supersonic
turbulence from CO overtone bands, but lower linewidths for H$_2$O in SVS13, suggesting strong turbulence in the upper
layers of the inner disk. On the other hand, for the outer disk of
\object{DM Tau}, \citet{Dartois+etal_2003} find marginally larger local velocity
dispersion in $^{13}$CO than in $^{12}$CO, suggesting lower levels
of turbulence in the upper layers than in the disk plane. The
derived broadening, 0.07 to 0.14 km.s$^{-1}$, is subsonic, but
its exact value is difficult to assess because of the contribution
of the thermal component. \citet{Pietu+etal_2007} find
similar results for \object{MWC 480} and \object{LkCa 15}. More recently, \citet{Hughes+etal_2011}
have obtained a low value for \object{TW Hya} ($< 0.04$ km.s$^{-1}$), but up to
0.3 km.s$^{-1}$ for the Herbig Ae star \object{HD 163296} from CO J=3-2 observations.

However, although $^{12}$CO is an interesting indicator, because of the significant
opacities of the lower lying transitions, it only probes
a thin upper layer in the disk, where strong
temperature gradients may affect the apparent linewidth. To build a complete
picture of turbulence in disks, it is thus essential to
constrain the turbulence from other (optically thinner) tracers. CO isotopologues and other molecules
like  HCO$^+$ and CN yield typical linewidths around 0.18 km.s$^{-1}$ at 300 AU,
but attributing this to thermal or nonthermal motions remains difficult,
as their thermal width at 30 K is 0.14 km.s$^{-1}$ \citep{Pietu+etal_2007, Chapillon+etal_2012}.
Moreover, current chemical models fail to explain their vertical location in the molecular
layer since CN, CCH or HCN appear to be in a colder molecular layer than predicted
\citep{Henning+etal_2010, Chapillon+etal_2012}.

It is then important to reduce the uncertainty (and possible bias) introduced by the
thermal contribution to the linewidth.
Among molecules detected in disks, only CS is both abundant enough and sufficiently
heavier than the previously cited ones.

We report here on the study of the CS molecule in the DM Tau disk.
Observations are presented in Section \ref{sec:obs}, disk properties are derived
in Section \ref{sec:ana}, and we discuss the results in Section \ref{sec:dis}.

\section{Observations}
\label{sec:obs}
Observations were carried out with the IRAM Plateau de Bure interferometer. We observed
two different transitions of CS, the J=3-2 line at 146.969047 GHz and the J=5-4 line
at 244.935609 GHz.

\begin{figure*}
   \includegraphics[width=18.0cm]{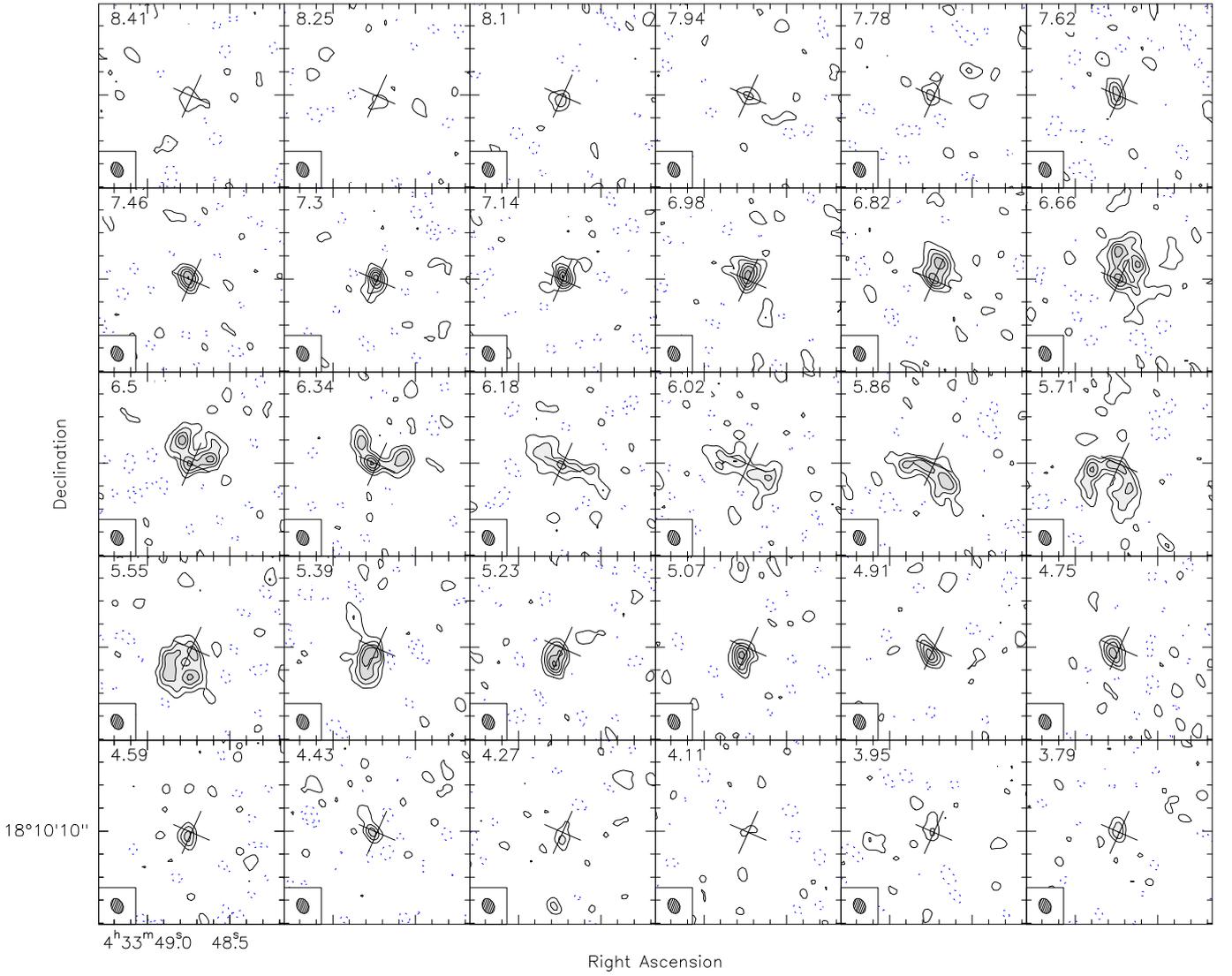}
 \caption{Channel maps of the CS J=3-2 emission towards DM Tau.
 The angular resolution is $1.4''\times 1.0''$ and the spectral resolution 0.16 km\,s$^{-1}$.
 Contour spacing is 10 mJy/beam, corresponding to $2 \sigma$ and 0.4 K brightness.
 The cross indicates the position, orientation,
 and aspect ratio of the dust disk.}
  \label{fig:cs32}
\end{figure*}

For the J=3-2 transition, we used the BC configuration (project RB8F).
The B configuration, with baselines
up to 408 m was observed on 18 Mar 2008, while the more compact C configuration was observed
on 15 Nov 2008. Phase noise ranged from $15^\circ$ up to $55^\circ$ on the longest baselines.
Amplitude calibration was done using nearby quasars and led to an rms level of about $5 \%$.
Absolute flux calibration was done using MWC\,349 as a reference.  The
angular resolution is  $1.4''\times 1.0''$. The correlator provides only a limited spectral resolution
with a channel spacing of 0.08 km.s$^{-1}$, but with an effective resolution of 1.59 times
the channel spacing (i.e. 0.126 km.s$^{-1}$) owing to Welch apodization of the Fourier spectrum.
At this resolution, the brightness sensitivity is 8.8 mJy/beam, which corresponds to 0.35 K at this angular
resolution. Channel maps of CS(3-2) obtained after smoothing in velocity
to 0.16 km.s$^{-1}$ are presented in Fig.\ref{fig:cs32}, but our analysis uses the
unsmoothed data.

\begin{figure}
   \includegraphics[width=7.0cm]{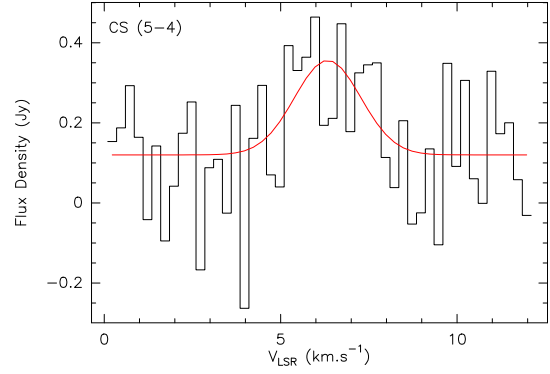}
 \caption{Spectrum of CS J=5-4 line towards DM Tau.
 The angular resolution is $3.5''$ and the spectral resolution 0.25 km\,s$^{-1}$.
 The curve is a best-fit Gaussian to emphasize the detection.}
  \label{fig:cs54}
\end{figure}

The CS(5-4) data were obtained on 11 Nov 2004, 17 Nov 2004, and 03 Apr 2005 (Project OA45).
The natural weighting spatial resolution is $2.6'' \times 1.8''$ at PA 85$^\circ$. We smoothed
the data to a spectral resolution of 0.25 km.s$^{-1}$ for the analysis.
The brightness sensitivity is 90 mJy/beam, corresponding to 0.4 K.
The CS(5-4) is barely detected (at the $5 \sigma$ level): a spectrum towards DM Tau obtained at
3.5$''$ resolution is shown in Fig.\ref{fig:cs54}. The data is compatible
with the integrated spectrum obtained with the 30-m by \citet{Dutrey+etal_1997}
and a source size of about 5$''$, with an integrated line flux $\simeq 0.6 \pm 0.1$ Jy.km.s$^{-1}$.

The relative calibration accuracy between the two frequencies is better than 10 \%.
It is based on the precise spectral index of 0.6 for the reference flux calibration source MWC\,349.

\section{Results}
\label{sec:ana}

\subsection{Principle of the method}
\begin{figure}[!t]
\begin{center}
\resizebox{6.0cm}{!}{\includegraphics[angle=270]{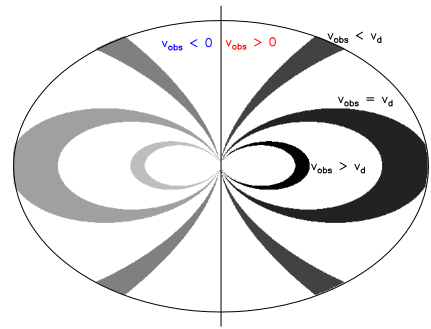}} \caption[Cinematique]{\label{fig:cinematique}
Regions of the disk that yield equal projected velocities $v_\mathrm{obs}$. The ellipse is the
projection of the disk's outer edge.}
\end{center}
\end{figure}

Line formation in a Keplerian disk is strongly constrained by the velocity gradient
\citep[see, e.g.][in the context of cataclysmic binaries]{Horne+Marsh_1986}.
The line of sight velocity
(in the system rest frame) is
\begin{equation}
V_\mathrm{obs}(r,\theta) = \sqrt{GM_{*}/r} \sin{i}\cos{\theta}
\end{equation}
where $r,\theta$ are the cylindrical coordinates in the disk plane.
The locii of isovelocity are given by
\begin{eqnarray}
r(\theta) &=& ( GM_{*}/V_\mathrm{obs}^2 ) \sin^2{i}\cos^2{\theta} .
\end{eqnarray}
With a finite local linewidth $\Delta v$ (assuming rectangular line shape for simplification), the line at
a given velocity $V_\mathrm{obs}$ originates in  a region included between $r_i(\theta)$ and $r_s(\theta)$ :
\begin{eqnarray}
r_i(\theta) &=& \frac{GM_{*}}{(V_\mathrm{obs}+\Delta v/2)^2}\sin^2{i}\cos^2{\theta} \\
r_s(\theta) &=& \min\left[R_\mathrm{out},\frac{GM_{*}}{(V_\mathrm{obs}-\Delta v/2)^2} .
\sin^2{i}\cos^2{\theta}\right]
\end{eqnarray}
Figure \ref{fig:cinematique} indicates the regions of equal projected velocities for six different values:
$V_\mathrm{obs}>v_d$, $V_\mathrm{obs} = v_d$, $V_\mathrm{obs}<v_d$, and their symmetric counterpart at negative
velocities, where $v_d$
\begin{eqnarray}
v_d &=&  \sqrt{GM_{*} / R_\mathrm{out}} \sin{i} \label{eq:vd}
\end{eqnarray}
is the projected velocity at the outer disk radius $R_\mathrm{out}$.
Figure \ref{fig:cinematique} and Eq.\ref{eq:vd} thus show that the basic shape of the emission as a function
of velocity depends only on inclination $i$, while the stellar mass gives the scaling
factor for the velocity scale. The intensity along the curves given by Eq.\ref{eq:vd} will
depend on the surface density $\Sigma(r)$ and temperature $T(r)$ profiles.  On the  other hand, the spatial spread
perpendicular to this nominal shape depends \textit{only} on the intrinsic linewidth $\Delta v$.
With sufficient angular resolution, we thus expect that $\Delta v$ is only very weakly coupled to any of the
other disk parameters. The angular resolution should be at least high enough to measure
the shape described by Eq.\ref{eq:vd}, i.e., measure the inclination $i$. A more stringent requirement
is to resolve the transverse size perpendicular to the iso-velocity curves. An insufficient
resolution would introduce a coupling between $\Delta v$ and
$\Sigma$ or $T$ as a smaller spatial spread can be compensated for by higher temperature (or opacity) to
lead to the same flux density.

\subsection{Simple Model}
\label{sec:ana:power}
To derive quantitative information from the data, we used the radiative transfer code adapted to disk DISKFIT
 \citep[see][for a detailed description]{Pietu+etal_2007}. In a first step (Model A), the CS column densities and rotational excitation
 temperature distributions are parameterized using a power law of the radius:
\begin{equation}
 \Sigma_\mathrm{CS} (r)=\Sigma_0\left(\frac{r}{R_s}\right)^{-p}
 \end{equation}
\begin{equation}
 T(r)=T_0\left(\frac{r}{R_t}\right)^{-q} .
 \end{equation}
As explained by \citet{Pietu+etal_2007}, $T(r)$ is the line excitation temperature, and it is assumed to apply to all
other levels to derive $\Sigma(r)$. We further assume LTE, so that $T(r)$ is now also the kinetic temperature. Then, the
local linewidth consists of a thermal component, plus a turbulent contribution
\begin{equation}
\Delta V(r) = \sqrt{ \frac{2 k T(r)}{\mu m_H} + \delta V_\mathrm{tu}(r)^2 }
\end{equation}
where $\mu = 44$ is the CS molecular weight and $m_H$ the atomic mass unit.
We note that $\Delta V(r)$ is a half width at $1/e$ and
that the full width at half maximum would be 1.66 times larger.
The turbulent component is also parameterized by a power law
\begin{equation}
\delta V_\mathrm{tu} (r) = dV_0 \left(\frac{r}{R_v}\right)^{-e_v} .
\label{eq:width}
\end{equation}
Finally, CS is assumed to be homogeneously distributed in the $z$ direction, with a scale
height given by the disk midplane temperature. The density is given by
\begin{equation}
n_\mathrm{CS} = \frac{\Sigma_\mathrm{CS}}{\sqrt{\pi} H(r)} \exp \left(-(z/H(r))^2\right)
\end{equation}
and the scale height is set to
\begin{equation}
H(r) = H_0 (r/R_h)^{1.25}
\end{equation}
with $H_0 = 16$ AU at $R_h = 100$ AU. This definition leads to a scale height that
is $\sqrt{2}$ times larger than the $H(r)=c_s/\Omega_K$ convention, where $c_s$ is the sound
speed and $\Omega_K$ the Keplerian angular velocity.

All the other geometric disk parameters (systemic velocity, inclination and position angle of the disk
axis) were taken from the analysis of the CO isotopologues performed by \citet{Pietu+etal_2007}.
We also checked that CS gave consistent results with these independent measurements and
that the errorbars on these geometric parameters did not affect the other derived parameters.
A summary of the disk characteristics is given in Table \ref{tab:disk}.

Synthetic images are generated using the radiative transfer code DISKFIT, and model visibilities
computed for each of the observed $u,v$ points. Minimization on the disk free parameters is then
made using a $\chi^2$ criterion based on the difference between the observed and simulated visibilities.
This process avoids the nonlinearities that result from deconvolution.
A modified Levenberg-Marquardt method is used to locate the minima, with multiple restarts
to avoid secondary minima.
The linewidths we are searching for are not very large compared to the effective
correlator resolution (0.126 km.s$^{-1}$, see Section \ref{sec:obs}). Thus, we
simulated the correlator spectral response by oversampling by a factor 4 in velocity space,
i.e., using a velocity sampling bin of 0.02 km.s$^{-1}$,
followed by a convolution with a kernel mimicking the Welch apodization. This kernel
extended over 12 consecutive channels in the oversampled frequency space.

Under these assumptions, the CS(3-2) line emission can be used to constrain
$\Sigma_0,p,T_0,q,dV_0$ and $e_v$. Errobars around the minimum are derived
from the covariance matrix. A proper choice of the pivot values $R_t, R_v, R_s$ must be made to minimize
the errors on the power laws \citep[see][for details]{Pietu+etal_2007}. With our angular resolution
and CS images, $R_t=R_v=R_s = 300$ AU is a good compromise.
A degeneracy occurs between $\Sigma_0$ and $T_0$ if the
line is optically thin (for example, in the high temperature limit, the emission only depends on $\Sigma_0/T_0$), but
this degeneracy is broken if the surface density profile is steep enough to produce an optically
thick core \citep[see][]{Dutrey+etal_2007}. Indeed, from a global fit,
we find $T(r) = (7.2 \pm 0.4) (r/300 \mathrm{AU})^{-0.63\pm0.09} \mathrm{~K}.$ 
 Assuming this temperature to be the kinetic temperature, we find
an intrinsic \textit{nonthermal} half width at 1/e,
$\delta V_t(r) = 0.13 \pm 0.03 (r/300 \mathrm{AU})^{-0.38 \pm 0.45}$ km.s$^{-1}$. 
The outer radius is found to be $540 \pm 10$ AU.

The derived excitation temperature is unlikely to differ much from the kinetic temperature, because
the densities in the disks are substantially higher than the CS(3-2) critical densities, even at the outer radius.
To explain the observed \textit{total} intrinsic width by
thermal motions ($\delta V_\mathrm{th} = \sqrt{2 k T(r) / \mu_\mathrm{CS} m_H} $, with
$\mu_\mathrm{CS} = 44$) would require a high value of $T \simeq 50$ K at 300 AU.
This simple analysis thus indicates that pure thermal motions cannot reproduce the observed emission pattern,
under the assumptions of Model A.

\begin{figure}[!t]
\begin{center}
\includegraphics[width=\columnwidth]{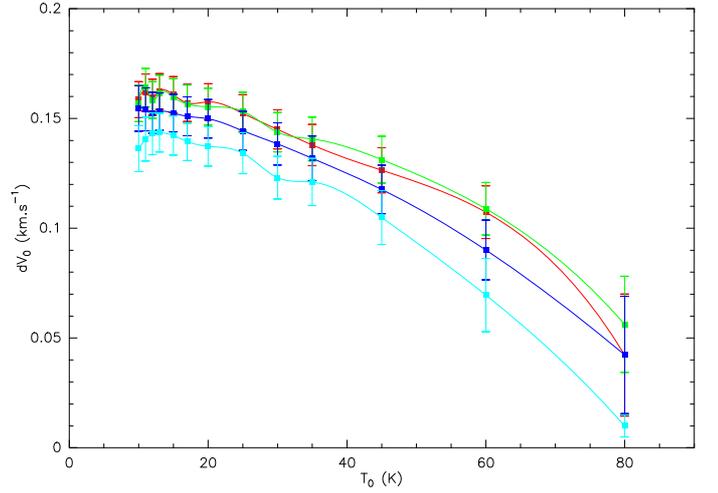} \caption[dvtk]{\label{fig:dvtk}
Derived nonthermal linewidth $dV_0$ at $R_v= 300$ AU as a function of assumed kinetic temperature
profile. $T_0$ is the temperature at $R_t = 300$ AU. The 4 curves correspond
to different exponents $q = 0$ (red), 0.2 (green), 0.4 (blue), and $0.6$ (cyan).
Errorbars are $\pm 1 \sigma$.}
\end{center}
\end{figure}
To strengthen this conclusion,
as our main goal is to constrain $\delta V_t(r)$, we also explored a much wider range
of possible temperature profiles, varying $T_0$ and $q$. Results for $dV_0$ as a function of assumed
$T_0$ are given in Fig.\ref{fig:dvtk} for four values of $q=0, 0.2, 0.4$, and $0.6$. In each case, because the exponent
$e_v$ is not well constrained, we assumed $e_v = q/2$, which corresponds to a constant Mach number.
Figure \ref{fig:dvtk} shows that, unless the kinetic temperature is very high, the nonthermal
linewidth must be $\simeq 0.10 - 0.15$ km.s$^{-1}$ to represent the observations.

\begin{figure}[!t]
\begin{center}
\includegraphics[width=\columnwidth]{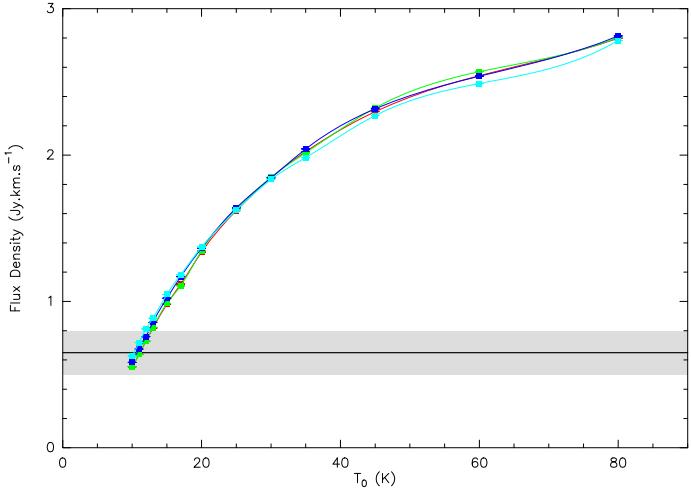} \caption[cs54tk]{\label{fig:cs54tk}
Predicted CS J=5-4 line flux from the best fit models derived from CS J=3-2 observations, as
a function of assumed kinetic temperature profile. The 4 curves corresponding to the different exponents q = 0,
0.2, 0.4, and 0.6 are essentially degenerate. Errorbars are $\pm 1 \sigma$.
The shaded gray area indicates the measured
line flux and its $\pm 1 \sigma$ range. Calibration uncertainty is not included.}
\end{center}
\end{figure}
The CS(3-2) data set itself
suggests low temperatures, but the surface density profile is essentially flat ($p = 0.14 \pm 0.20$), so the constraint
may be considered as relatively weak.  The temperature found  with CS(3-2) is quite compatible
with the one determined from CCH \citep{Henning+etal_2010} and CN and HCN \citep{Chapillon+etal_2012}.
Moreover, the CS(5-4) data, although quite noisy, can be used to rule out high
temperatures. Figure \ref{fig:cs54tk} shows the integrated line flux for the CS(5-4) data using the
best fit parameters found from CS(3-2) as function of $T_0$ and $q$. Clearly,
high temperatures ($> 20$ K) can be ruled out, since they would produce a
much brighter CS(5-4) emission than observed. This is consistent
with the temperature derived from other tracers, although subthermal excitation of the CS(5-4)
in a warmer medium cannot be ruled out.\footnote{This analysis also implicitly assumes a lack of vertical
(excitation) temperature gradient. On one hand, subthermal excitation is less unlikely for the J=5-4
line than for the J=3-2, but on the other,  the J=5-4 has higher opacities, and can thus probe
higher, warmer levels. The balance between the two effects cannot be estimated without
detailed non-LTE radiative transfer models.}

Low temperatures are thus strongly suggested, although not firmly proven, by
this study. In any case, the uncertainties on temperature profile have a negligible
influence on the key parameters of this study, $\delta V_0$, as well as on $p_\mathrm{CS}$.
However, they can affect the derived CS column density by a factor 1.5-2.

\subsection{A more realistic approach ?}
\label{sec:ana:viscous}
The assumption of homogeneous vertical mixing (constant abundance as a function of $z$) is a strong prior.
In practice, chemical models predict that molecules are strongly depleted in the disk plane, owing to the
condensation of volatiles on dust grains.  Thus, the emission of molecules is expected to come from one or
two scale heights above the disk plane. As the scale height is typically 10 \% of the radius, this means the
emission comes from regions inclined $\pm \delta i \simeq 6^\circ$ from the disk midplane.
Because the line of sight velocity is proportional to $\sin (i)$, these two different inclinations result in
a broadening of the projected linewidth, which is not properly reproduced by the assumption of no vertical
abundance gradient. This is essentially as if the emission came from two disks, one inclined by $i+\delta i$
and the other  by $i - \delta i$ along the line of sight \citep[see][]{Semenov+etal_2008}.
To first order, the extra broadening due
to these inclination differences would then be proportional to the rotation velocity, i.e. would decrease as $\sqrt(r)$,
if $\delta i$ is constant. The exponent of the nonthermal linewidth law $e_v$ (see Eq.\ref{eq:width})
derived above is indeed quite close to the Keplerian exponent 0.5 (but with little significance).

In practice, $\delta i$ will vary with radius. On one hand, $h(r)/r$ increases slowly with $r$ (as $\approx r^{0.25}$).
On the other, the height above which molecules are found is expected to decrease in the outer parts of the disk,
because (interstellar or scattered) UV photons can penetrate deeper towards the disk plane. This leads to enhanced
photodesorption, and brings the molecular layer closer to the disk plane. As a consequence, the resulting increase
in inclination spread is smaller in the outer regions than closer to the star. This could also contribute to the
apparent decrease in the nonthermal broadening found previously by our simple analysis.

It is thus essential to check whether nonthermal motions are also required to explain the observations
when CS is confined above the disk midplane.  Although sophisticated chemical models exist, they require a prior
knowledge of the H$_2$ density structure. This density structure is still very uncertain, and parametric chemical
models (which would explore a range of possible solutions) do not yet exist in this respect.

We thus adopted a second parametrization of the CS emission from the disk,  Model B, which
differs from Model A in two ways.  The CS surface density
is no longer a power law.  Instead, the disk surface density (dominated by
molecular hydrogen) follows an exponentially tapered law
\begin{equation}\label{eq:edge}
\Sigma_H(r) = \Sigma_0 \left(\frac{r}{R_0}\right)^{-\gamma}  \exp\left(-(r/R_c)^{2-\gamma}\right) ,
\end{equation}
which is the self-similar solution of the disk evolution when the viscosity is a (time constant) power
law of radius \citep{LindenBell+Pringle_1974}.
CS is then assumed to only be present when the hydrogen column density from the current point
towards the disk exterior (perpendicular to the disk midplane) is lower than some given
threshold, $\Sigma_d$, the \textit{depletion} threshold. In regions where CS exists, we
further assume that its abundance is only a power law of radius.
The disk is assumed to be vertically isothermal. The column density above a given
height $z$ is then given by
\begin{equation}
\Sigma_\infty(r,z) = \int_z^\infty n(r,u) du = \frac{\Sigma(r)}{2} \left(1 - \mathrm{erf} \left(\frac{z}{H(r)}\right) \right) ,
\end{equation}
from which the depletion height $z_d(r)$ can be recovered as a function of $\Sigma_d/\Sigma(r)$
\begin{equation}
z_d(r) = H(r) \, \mathrm{erf}^{-1}\left(1-\frac{2 \Sigma_d}{\Sigma(r)}\right) .
\end{equation}
The CS abundance is therefore given by
\begin{eqnarray}
 X_\mathrm{CS}(r) = & 0 & \mathrm{for~~~~} z < z_d(r) \\
 X_\mathrm{CS}(r) = & X_\mathrm{CS}^0 (r/R_0)^{-p_\mathrm{CS}} & \mathrm{for~~~~} z > z_d(r) .
\end{eqnarray}
In this model, the overall surface density of CS thus follows the law
\begin{equation}
 \Sigma_\mathrm{CS}(r) = X_\mathrm{CS}^0 \left(\frac{r}{R_0}\right)^{-p_\mathrm{CS}} min(\Sigma_d, \Sigma_H(r)) .
\end{equation}

In the modeling, $R_c$ and $\gamma$ are fixed parameters. We used $R_c = 180$ AU and $\gamma = 0.55$,
which were found by \citet{Guilloteau+etal_2011} from high-resolution, dual-frequency continuum, and
$\Sigma_0$ is also taken from this article to obtain  a disk mass of
$0.03 \Msun$, which gives $\Sigma_0(100 \mathrm{AU}) = 9.6\,10^{23}$ cm$^{-2}$ (H$_2$ molecules).

Although imposing the density structure could allow non-LTE modeling to be performed, considering
the large uncertainties on the real density structure\footnote{As stated before, the dust density
distribution in disks (and a fortiori in the
DM Tau disk) is poorly known. For example, the adopted values do not consider the variations in the
dust properties with radius found by \citet{Guilloteau+etal_2011}. \citet{Andrews+etal_2012}
also argue that no simple model represents both the gas and dust distributions in TW Hya.},
CS is assumed to be thermalized because densities are in general high enough, even at one to two scale heights.
The remaining free parameters are now $X_\mathrm{CS}^0$, $p_\mathrm{CS}$,
$T_0$, $q$, $R_\mathrm{out}$, the turbulent width $dV_0$ ($e_V$ being poorly constrained, we used
$e_V = q/2$, a constant Mach number), and the depletion threshold $\Sigma_d$.

Although the rationale for the choice of parameters is based on behaviors predicted by chemical models,
it is important to stress that our approach is still purely parametric. In particular, we make no \textit{a priori} link
between the thermal structure of the disk and the location of molecules.

The optimal value of $\Sigma_d$ was first located through a grid of best-fit solutions computed with $\delta V$
and $\Sigma_d$ and $T_0$ as fixed parameters, and $X_\mathrm{CS}^0$, $p_\mathrm{CS}$ and $R_\mathrm{out}$ as
free parameters. The temperature was set to the best-fit solution found for the power law; similar results were
obtained when allowing the temperature law to vary. This grid search did not
reveal any secondary minimum. We then started from the best fit found in this grid search to proceed with a
global fit
of the parameters, including the temperature law.
The best fit results are presented in Table \ref{tab:disk}. This model provides a slightly better fit to the
CS(3-2) data than the power-law case.

\subsubsection{Turbulent width}
We find a nonthermal component slightly smaller than that derived in the power-law case, but still significant
$dV_0= 0.11\pm 0.015$ km.s$^{-1}$ for $H_0 = 16$ AU.
A lower value is expected, since part of the spatial extent at a given velocity is due to the
spread in inclinations, which is larger in Model B than in Model A (which assumed vertically uniform CS abundances).

\subsubsection{Depletion column density}
The best fit ``depletion'' column density is  $\Sigma_d = 10^{21.7\pm0.1}$ cm$^{-2}$. However,
interpretation of this value is not straightforward. Changing the scale height from 16 AU
(at 100 AU) to 8 or 32 AU does not significantly affect the derived $\Sigma_d$. This indicates that
$\Sigma_d$ is not constrained by the geometry alone. In fact, this value is found because
the CS(3-2) images are apparently better represented by a simple power law for the CS surface density than
by the more complex radial profiles that can  result from the hypotheses made in Model B.  With $\Sigma_d
 \simeq 10^{22}$  cm$^{-2}$ and a power law for the CS abundance, the CS surface density behaves
as a power law (of exponent $p = 0.4 \pm 0.2$) out to $R \approx 400$ AU, close enough to the outer radius of the
CS distribution. With $H_0 = 16$ AU,
model B also leads to slightly higher (but less constrained) temperatures than Model A, around $10\pm2$ K at 300 AU.

\subsubsection{Vertical location}
\begin{figure}
   \includegraphics[width=\columnwidth]{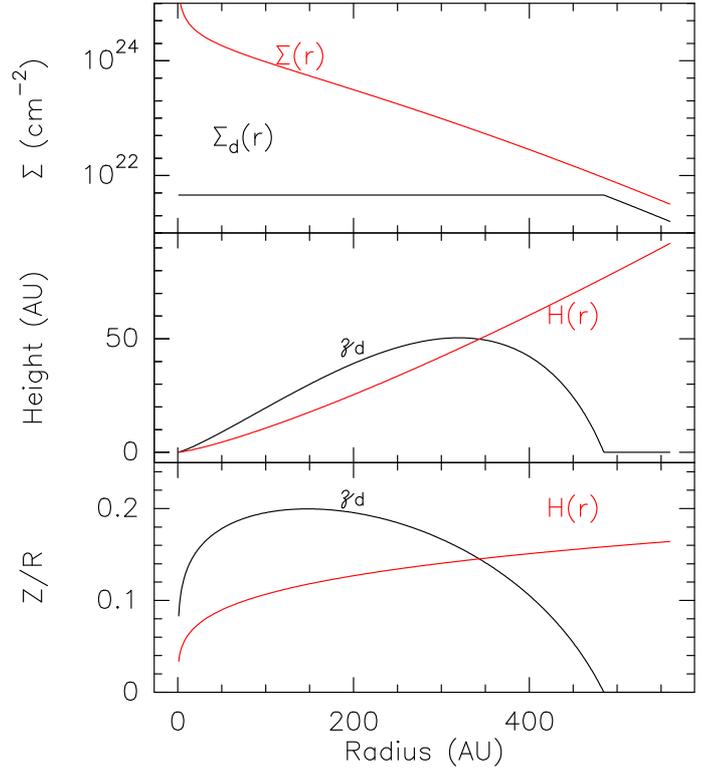}
 \caption{Top: Total
 disk surface
 density $\Sigma(r)$ and column density $\Sigma_d(r)$ above the depletion height
 $z_d(r)$. Middle: Depletion height $z_d(r)$ as a function of radius in the best
 fit model.  The scale height $H(r)$ is also indicated (in red). Bottom: as
 above, but for $z_d(r)/r$ and $H(r)/r$.}
  \label{fig:deplete}
\end{figure}

Although $\Sigma_d$ is mostly constrained by the radial brightness profile in this
case, it does not imply that the vertical distribution is not constrained. Starting from
the best fit of the radial CS distribution, we can actually get an estimate of the system geometry by
treating the scale height as a free parameter. We find $H_0 = 9 \pm 1.5$ (see Table \ref{tab:disk}),
which is consistent with the temperature derived from CS, which indicates $H_0 = 10.4 \pm 1.2$.
The required turbulent width increases when the scale height is diminished.
The depletion height $z_d$ derived from this best fit is plotted in Fig.\ref{fig:deplete}, together
with the corresponding value of $\Sigma_d$. It is important to note that the shape of $z_d$ is
prescribed by our model assumptions. In reality, the distribution of molecules can be substantially
different, but our analysis nevertheless indicates that a substantial fraction of the CS molecules
must reside around one scale height.

\subsection{Comparison with chemical models}
\label{sec:ana:chem}

A chemical model for sulfur bearing molecules was developed by \citet{Dutrey+etal_2011}
to explain the integrated line profiles obtained with the IRAM 30-m.
The model predictions were done with the NAUTILUS code \citep{Hersant+etal_2009},
which computes the chemical composition of the gas and interstellar ices
as a function of time. Bimolecular reactions between gas-phase species and
interactions with direct UV photons and cosmic-ray particles, as well as UV
photons induced by cosmic-rays,  are considered, with a network of 4406
reactions for 460 species. Interactions with grain surfaces (including direct
and UV induced dissociations), as well as reactions at the surfaces, through
1773 reactions, determine the abundance of 195 surface species. The NAUTILUS
code is run in the vertical 1D approximation at different radii of the disk
(from 50 to 500 AU). More details on the model parameters and the physical
structure of the disk can be found in \citet{Dutrey+etal_2011}. The
observations of CS presented in this paper have been compared to the Case C for
DM Tau of \citet[][see their Tables 4 and 5]{Dutrey+etal_2011}.

\begin{figure}
   \includegraphics[width=\columnwidth]{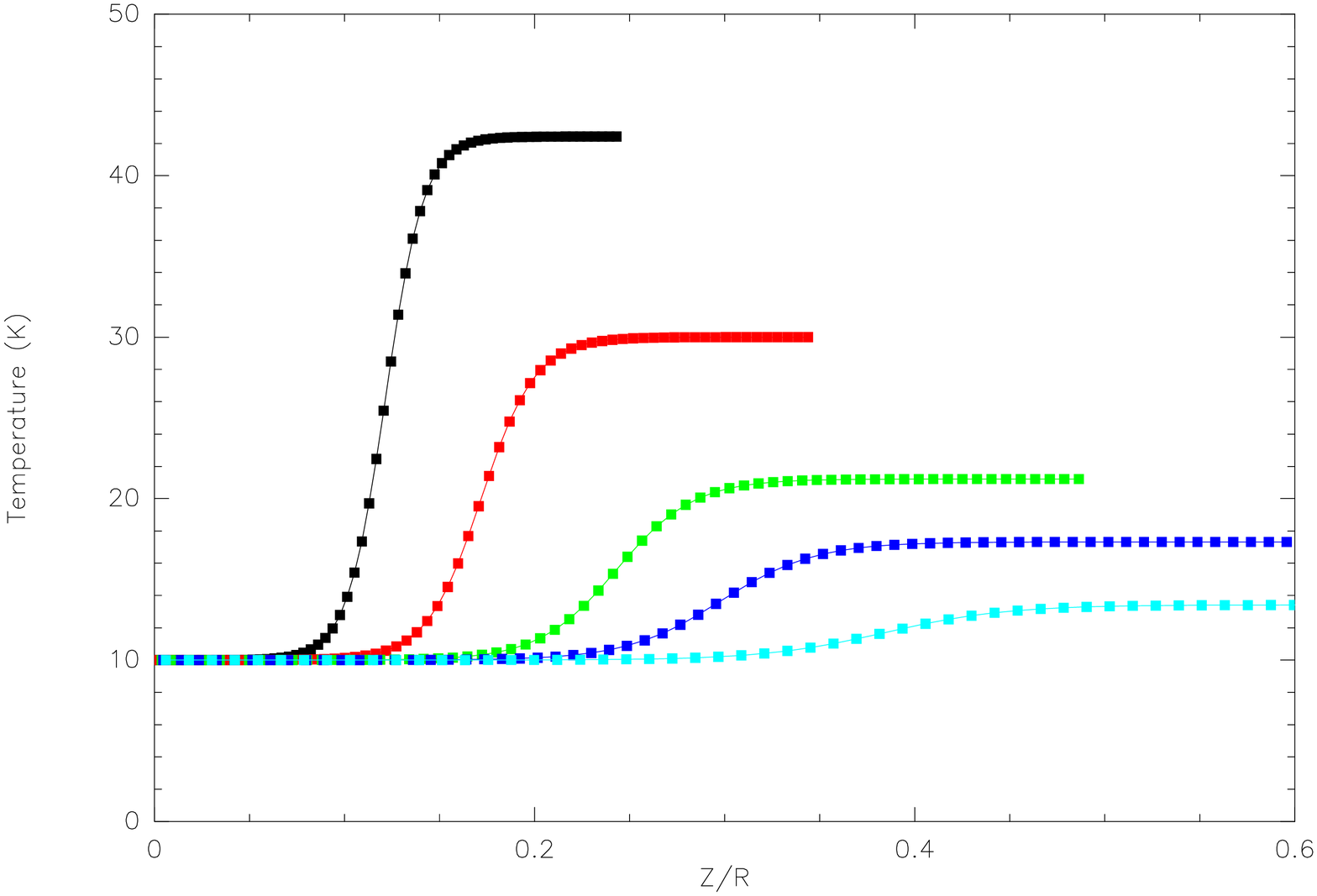}
   \includegraphics[width=\columnwidth]{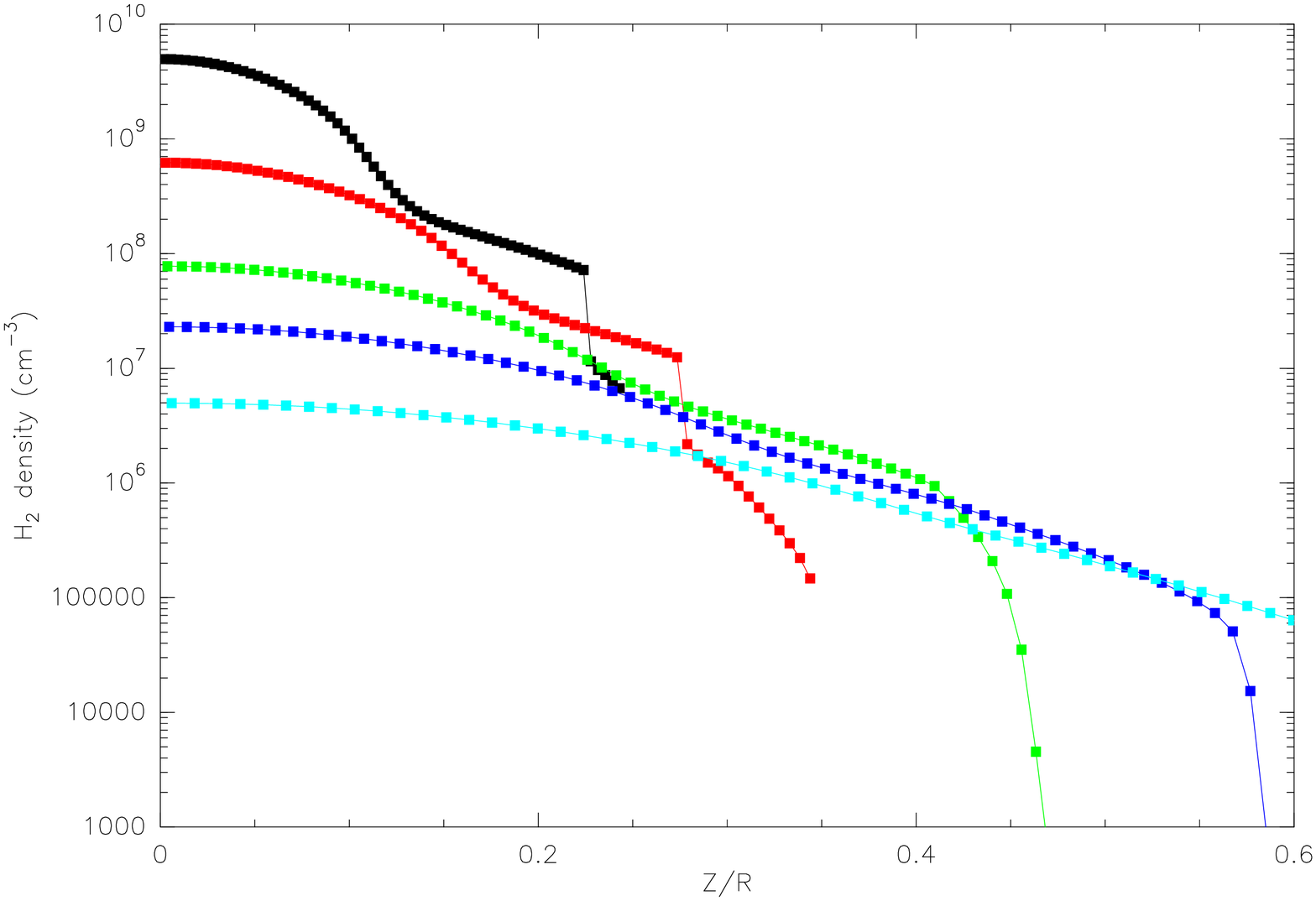}
   \includegraphics[width=\columnwidth]{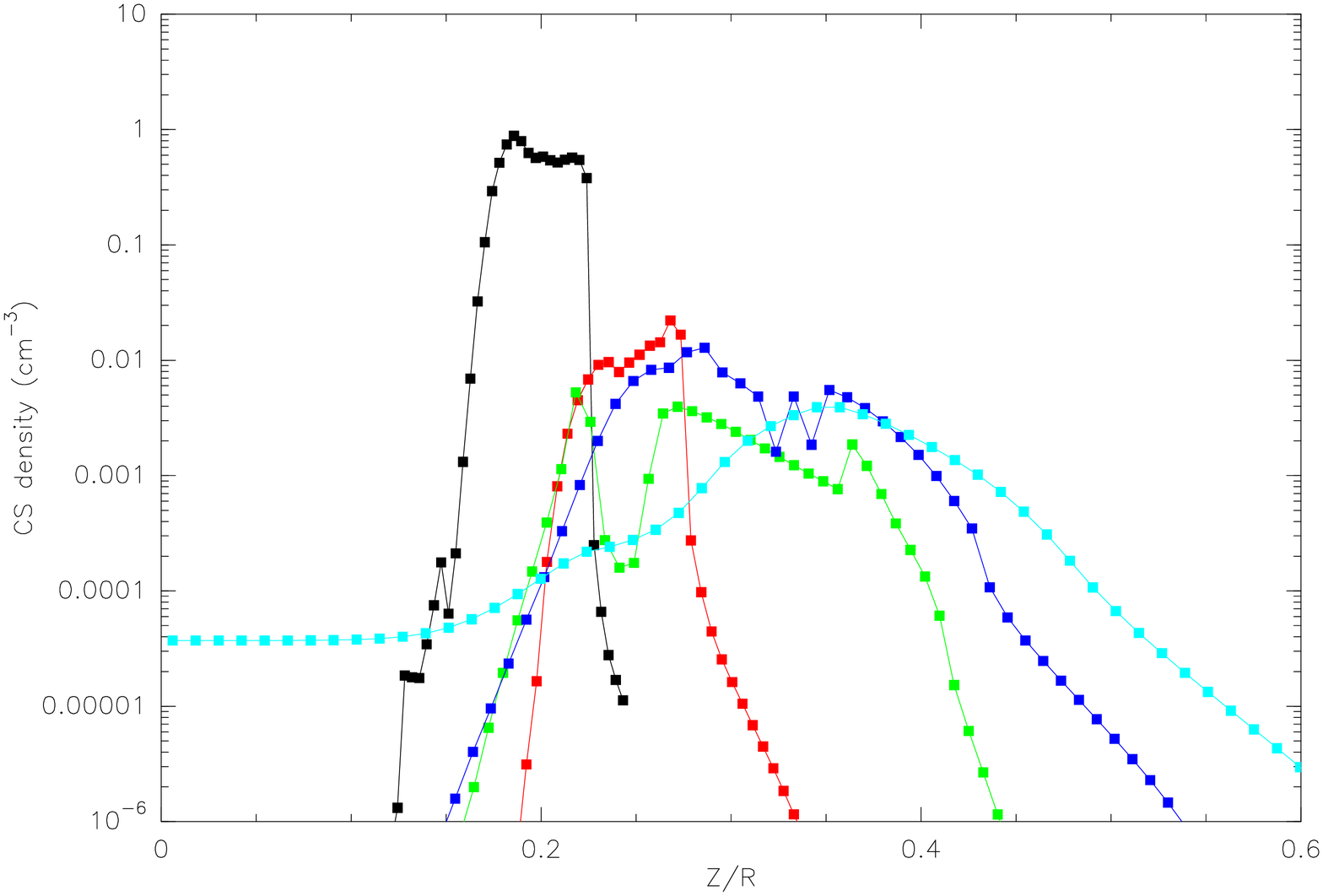}
 \caption{Predicted distribution of Temperature (top), H$_2$ (middle) and CS (bottom) molecule densities as a function
 of height for several radii: 50 (black), 100 (red), 200 (green),
 300 (blue), and 500 AU (magenta). }
  \label{fig:cs-z}
\end{figure}

Although this model does not use the same underlying density structure, it is interesting
to compare its predictions to the results derived from the Plateau de Bure data.
Figure \ref{fig:cs-z} shows the distribution of CS molecules and the H$_2$ density and temperature
vertical profiles for several radii.  The range
of heights at which CS is found corresponds to an H$_2$ column density towards
the outside, $\Sigma_\infty(r,z)$ in the range of $10^{22} - 10^{23}$ cm$^{-2}$, except at 50 AU where the
warmer interior leads to enhanced CS abundance at larger depths.

The temperature is not the key parameter defining the molecular-rich layer. CS exists
at temperatures well below its evaporation temperature (30 K), as can be seen in
Fig.\ref{fig:cs-z}. The penetration
of UV radiation (scattered from the star and coming from the ISM) is more important in
defining this layer, which is limited by photodesorption near the disk plane and photodestruction
of all molecules near the disk surface. Temperature effects become significant only above 30 K, for which
high CS densities can be reached because of thermal desorption.

The enhanced CS surface density within 50 AU from the star is not probed by our observations, which are mostly
sensitive to the 100 - 500 AU region.  In this region, the predicted column density
increases slightly with radius \citep[see][Fig.4]{Dutrey+etal_2011}, while the analysis of our observations
indicates a roughly constant or slightly
decreasing column density.
The chemical model does not include any radial
or vertical mixing, that would be a natural consequence of the observed turbulent velocities.
Vertical mixing will on average decrease the altitude at which CS molecules
are found, while radial mixing will smooth out the CS \textit{abundance} distribution.
Consequently, radial mixing would tend to make the slope of the CS surface
density profile closer to the one of H$_2$ that is expected to vary roughly as $1/r$.
This smoothing effect due to mixing can be seen in the study of \citet[][see their Fig.11]{Semenov+Wiebe_2011},
who investigated the impact of radial and vertical turbulent mixing. \citet{Semenov+Wiebe_2011} also
find that the CS content can be increased by an order of magnitude by mixing. Higher angular resolutions
(0.1 -- 0.3$''$) with ALMA are required to directly probe the inner disk region and prove (or disprove) our theoretical
prediction of high CS abundance here.

\begin{figure}
   \includegraphics[width=\columnwidth]{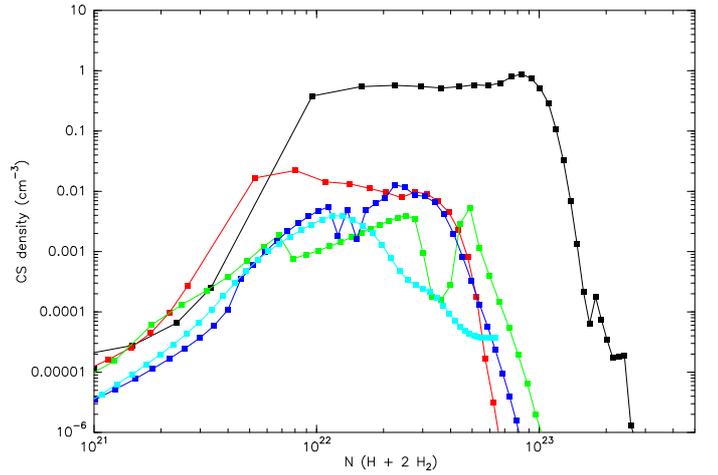}
 \caption{CS molecule density as a function of (essentially H$_2$) column density for several radii:
 50 (black), 100 (red), 200 (green), 300 (blue), and 500 AU (magenta). }
  \label{fig:cs-sigma}
\end{figure}

The chemical models predict that CS molecules are located between $z/r = 0.2$ and 0.4 (see Fig.\ref{fig:cs-z}), while
our best fit solution indicates $z_d/r \approx 0.15 - 0.2$. When expressed as a function
of $\Sigma_d$, the model predicts the range where CS is abundant reasonably well, a few $10^{22}$ to $\approx
10^{23}$ cm$^{-2}$ as shown by Fig.\ref{fig:cs-sigma}. The small discrepancies suggest that better agreement between
the chemical model and the observations can be obtained by assuming larger dust grains and a geometrically thinner
(i.e. colder) disk in the model. In this case, the molecular layer is expected to become closer to the disk plane
\cite[e.g.][]{Aikawa+Nomura_2006}, because of enhanced UV penetration and lower sticking efficiency.
Dust settling would also bring the molecular layer in the same direction, by minimizing the sticking efficiency
above the disk plane, but should be less important than grain growth since it only concerns the largest grains
that offer less surface area per unit mass.

\section{Discussion}
\label{sec:dis}

The analysis presented above unambiguously  indicates that a substantial additional line
broadening is required besides the thermal component and that the derived
magnitude of the turbulent motions is not too critically dependent on the assumptions
about the location of the CS molecules.

In the simple power-law approach (Model A), using our temperature measurement, a mean molecular
weight $\mu = 2.30$, the derived sound speed is
$ C_s(r) = \sqrt{k T(r)/\mu m_H} = 0.16 (r/300 \mathrm{AU})^{-0.3}\,\mathrm{km.s}^{-1}$
(using the temperature derived from CS for the kinetic
temperature), which corresponds to a Mach number $M = \delta V_t/C_s/\sqrt{2}$ around 0.5 and
a slightly lower value for Model B.

This result contrasts with the expectations from the $\alpha$ prescription
of the viscosity, where the turbulent width is related to $\alpha$ in a way that
depends on the nature of the turbulence. The value of
$\delta V$ ranges between a few $\alpha C_s$ to  $\approx  \sqrt{\alpha} C_s$
\citep{Cuzzi+etal_2001}. All indirect measurements of the $\alpha$ parameter \citep{Hartmann+etal_1998,
Hueso+Guillot_2005, Guilloteau+etal_2011} indicate $\alpha$ values between a few $10^{-4}$
and about 0.01, thus predicting much smaller turbulent widths, of at most
0.04 km.s$^{-1}$ if extrapolated near 300 AU.  However, these measurements often sample smaller
regions. \citet{Hartmann+etal_1998} derive these from accretion rates, and are thus sensitive
to the inner AUs only. The results of \citet{Hueso+Guillot_2005} and \citet{Guilloteau+etal_2011}
are based on viscous spreading of the dust disk, and are mostly sensitive to the 100 AUs range.

\citet{Hughes+etal_2011}
obtained a low value for TW Hya ($< 0.04$ km.s$^{-1}$), but up to
0.3 km.s$^{-1}$ for the Herbig Ae star HD\,163296 from CO J=3-2 observations.
An attractive explanation of these very different turbulence levels could have been
the age difference between these two stars,  with turbulence level decreasing with age as independently suggested
by the \citet{Guilloteau+etal_2011} results. Our DM Tau result does not fit well in this scheme, as DM Tau is
rather old, too. Differences in inclinations between the nearly face-on disk of TW Hya and the more inclined ones
of DM\,Tau and HD\,163296 could also play a role if the turbulence had significant anisotropy. However, shearing box
simulations of MRI induced turbulence by  \citet{Simon+Armitage+Beckwith_2011} show no evidence for such anisotropy.
On the other hand, \citet{Flock+etal_2011} and \citet{Fromang+Nelson_2006} find small anisotropy,
the radial velocity fluctuations being about 1.5 -- 2 times larger than in the azimuthal or vertical directions.

It is tempting to ascribe the difference between the constraints on $\alpha$ (sampling the disk plane)
and our measured turbulence level (sampling 1-2 scale heights) to the stratification of turbulence predicted by
MHD models. \citet{Fromang+Nelson_2006} and \citet{Flock+etal_2011} show that the velocity dispersion increases
quickly with height above two scale heights (with the $c_s/\Omega_K$ definition, i.e. at $\sqrt{2} H(r)$ with our definition).
However, they only reach a Mach number of 0.4 at four scale heights ($3 H(r)$), which appears insufficient to explain
our result.

The discrepancy between the apparent linewidths and the estimates of the viscosity may also point towards
a weakness in our theoretical understanding of the turbulence. The turbulent width measures velocity dispersion,
while the $\alpha$ viscosity measures a transport coefficient. They are
related through the effective mixing length \citep{Prandtl_1925}, which may not be a simple
function of $\alpha$ and $H$, as is usually assumed in the $\alpha$ model. In
other words, the aforementioned relation between the turbulent Mach number and
$\alpha$ may reach its limits in the disk outer parts.
It is worth pointing out that the TW Hya disk is small (outer radius 230 AU), while those of HD\,163296 and DM Tau are
large ($> 600$ AU and 900 AU, respectively). The smaller turbulent width found in TW Hya is consistent with the
idea that viscous spreading is the main driver for the disk size. Measurements of the nonthermal linewidths for
a wider sample of objects with
different disk radii would be important in establishing this potential link.

It is also important to realize that, in all the analysis of the molecular line emission, the so-called
``turbulent'' width is an adjustable parameter that is used to fit deviations from the simple Keplerian
disk model with purely thermal line broadening. Departure from these ideal conditions will affect the
derived ``turbulent'' width. For example, we have explored in Sect.\ref{sec:ana} the impact of the location
of molecules above the disk plane, but other effects can exist. Small deviations from Keplerian rotation,
which are ignored in our analysis, would have no impact on the linewidth for a nearly face-on object like
TW Hya, but could affect more inclined objects like DM Tau ($i=35^\circ$) or HD163296 ($i=45^\circ$). Our
study of the depletion depth in Sect.\ref{sec:ana} suggests that disk warps would, however, need to be quite
significant ($\sim 10^\circ$) before affecting the required turbulent component.

\begin{table}
\caption{Disk parameters and fit results}
\begin{tabular}{lccl}
\hline
\hline
Geometric        & {Adopted} & Fitted  \\
Parameter        & Value     & Value from CS \\
\hline
Distance (pc)    &  140  & \\
PA  ($^\circ$)   & 65    & $65 \pm 2$   \\
$i$ ($^\circ$)   & -35   & $-35 \pm 1$  \\
$V_\mathrm{LSR}$ & 6.08  & $6.08 \pm 0.02$  \\
$V_{100}$ $(\dagger)$        & 2.16  & $2.17 \pm 0.10$ \\
$M_*$ ($\Msun$)    & 0.54  & $0.54 \pm 0.04$  \\
h                & -1.25 &   \\
\hline
Fitted   & \multicolumn{2}{c}{Density Model} \\
Value    & (A) Power Law   &  (B) Tapered Edge & Note\\
\hline
$\chi^2$ & 2468353 & 2468336 \\
\hline
$H_0$ (AU) (a)  & [16] & $9\pm1.5$ & (1)  \\
$T_0$ (K) (b)   & $7.2 \pm 0.4$ &  $8.0 \pm 1.3$  &  \\
$q$  & $0.63 \pm 0.09$ & $0.60 \pm 0.20$ & \\
$\Sigma_\mathrm{CS}$ (cm$^{-2}$) (b) & $5.9 \pm 2.5\,10^{12}$  & -  & (2) \\
$X_{CS}$ (b)    & -  & $4.2 \pm 4.8\,10^{-10}$ & (2) \\
$p_\mathrm{CS}$ & $0.13 \pm 0.20$ & $0.39 \pm 0.18$ &  \\
$\Sigma_d$ (cm$^{-2}$)  &  - & $\approx 10^{21.7 \pm 0.1}$ & (3) \\
$R_\mathrm{out}$ (AU) & $540 \pm 10$ & $> 580$ & \\
$dV_0$ (km.s$^{-1}$) (b) & $0.13 \pm 0.03$ & $0.12 \pm 0.025$ & \\
$e_V$ & $0.38 \pm 0.45$ & [0.3] & (1) \\
\hline
\end{tabular}
\tablefoot{($\dagger$) Rotation velocity (km.s$^{-1}$) at 100 AU, which determines
the stellar mass $M_*$. (a) at 100 AU, (b) at 300 AU. (1) a number between brackets [] indicate a fixed
parameter. (2) Large errorbar due to strong coupling with temperature.
(3) Error bar not symmetric; derivation from covariance matrix inaccurate.}
\label{tab:disk}
\end{table}

\section{Summary}
\label{sec:sum}

Using the IRAM Plateau de Bure interferometer, we imaged the CS (3-2) emission from the disk of DM Tau
at $\sim 1''$ resolution. Although noisy, CS(5-4) data obtained with the 30-m and at 2.5$''$ resolution from
the PdBI preclude temperatures exceeding about 20-30 K. The morphology of the emission indicates a significant
\textit{local} linewidth,  around 0.14 km\,s$^{-1}$ near 300 AU. This result appears robust with respect to
the assumed location of CS molecules above the disk's midplane. While the magnitude of this linewidth is in
agreement with measurements previously obtained by \citet{Pietu+etal_2007} and \citet{Chapillon+etal_2012} from
other molecules like HCO$^+$, CN, HCN or CO isotopologues, the relatively large molecular
weight of CS (44), combined with the temperature estimate and an accurate handling of the spectral
response of the correlator, leaves no doubt here that this local linewidth is
essentially due to a nonthermal (turbulent) component. The magnitude of these turbulent motions corresponds
to a Mach number around 0.3 -- 0.5.

Our data confirm the relatively low temperatures derived from other molecules  such as CN, HCN, and CCH.
\citep{Chapillon+etal_2012,Henning+etal_2010}. They also suggest the CS molecules are located above one
scale height from the
disk plane, in reasonable agreement with predictions from chemical models.

This work shows the potential of CS as a turbulence tracer. No other molecules detected so far
\citep[including the recently reported HC$_3$N,][]{Chapillon+etal_2012b} is both heavy
and abundant enough to provide a more sensitive tracer.  Also, CS offers a number of transitions in
the mm / submm domain, which probe critical densities ranging from a few $10^5$ to several $10^8$ cm$^{-3}$. This,
combined with its expected location in the molecular-rich layer, make this molecule a key probe of physical
conditions in disks. The constraint obtained for the geometric location of CS molecules with respect
to the disk plane, although still difficult to interpret, clearly indicates the potential of ALMA for probing
different depths.

\begin{acknowledgements}
We thank the Plateau de Bure staff for performing the observations.
This work was supported by the National Program PCMI from INSU-CNRS.
DS acknowledges support by the \textit{Deutsche Forschungsgemeinschaft} through
SPP~1385: ``The first ten million years of the solar system - a
planetary materials approach'' (SE 1962/1-2).

\end{acknowledgements}

\bibliography{cs-paper}

\begin{thebibliography}{43}
\expandafter\ifx\csname natexlab\endcsname\relax\def\natexlab#1{#1}\fi

\bibitem[{{Aikawa}(2007)}]{Aikawa_2007}
{Aikawa}, Y. 2007, \apjl, 656, L93

\bibitem[{{Aikawa} \& {Nomura}(2006)}]{Aikawa+Nomura_2006}
{Aikawa}, Y. \& {Nomura}, H. 2006, \apj, 642, 1152

\bibitem[{{Andrews} {et~al.}(2012){Andrews}, {Wilner}, {Hughes}, {Qi},
  {Rosenfeld}, {{\"O}berg}, {Birnstiel}, {Espaillat}, {Cieza}, {Williams},
  {Lin}, \& {Ho}}]{Andrews+etal_2012}
{Andrews}, S.~M., {Wilner}, D.~J., {Hughes}, A.~M., {et~al.} 2012, \apj, 744,
  162

\bibitem[{{Balbus} \& {Hawley}(1991)}]{Balbus+Hawley_1991}
{Balbus}, S.~A. \& {Hawley}, J.~F. 1991, \apj, 376, 214

\bibitem[{{Beckwith} {et~al.}(2000){Beckwith}, {Henning}, \&
  {Nakagawa}}]{Beckwith+etal_2000}
{Beckwith}, S.~V.~W., {Henning}, T., \& {Nakagawa}, Y. 2000, Protostars and
  Planets IV, 533

\bibitem[{{Carr} {et~al.}(2004){Carr}, {Tokunaga}, \&
  {Najita}}]{Carr+etal_2004}
{Carr}, J.~S., {Tokunaga}, A.~T., \& {Najita}, J. 2004, \apj, 603, 213

\bibitem[{{Chapillon} {et~al.}(2012{\natexlab{a}}){Chapillon}, {Dutrey},
  {Guilloteau}, {Pi{\'e}tu}, {Wakelam}, {Hersant}, {Gueth}, {Henning},
  {Launhardt}, {Schreyer}, \& {Semenov}}]{Chapillon+etal_2012b}
{Chapillon}, E., {Dutrey}, A., {Guilloteau}, S., {et~al.} 2012{\natexlab{a}},
  \apj, 756, 58

\bibitem[{{Chapillon} {et~al.}(2012{\natexlab{b}}){Chapillon}, {Guilloteau},
  {Dutrey}, {Pi{\'e}tu}, \& {Gu{\'e}lin}}]{Chapillon+etal_2012}
{Chapillon}, E., {Guilloteau}, S., {Dutrey}, A., {Pi{\'e}tu}, V., \&
  {Gu{\'e}lin}, M. 2012{\natexlab{b}}, \aap, 537, A60

\bibitem[{{Cuzzi} {et~al.}(2001){Cuzzi}, {Hogan}, {Paque}, \&
  {Dobrovolskis}}]{Cuzzi+etal_2001}
{Cuzzi}, J.~N., {Hogan}, R.~C., {Paque}, J.~M., \& {Dobrovolskis}, A.~R. 2001,
  \apj, 546, 496

\bibitem[{{Dartois} {et~al.}(2003){Dartois}, {Dutrey}, \&
  {Guilloteau}}]{Dartois+etal_2003}
{Dartois}, E., {Dutrey}, A., \& {Guilloteau}, S. 2003, \aap, 399, 773

\bibitem[{{Dutrey} {et~al.}(1997){Dutrey}, {Guilloteau}, \&
  {Guelin}}]{Dutrey+etal_1997}
{Dutrey}, A., {Guilloteau}, S., \& {Guelin}, M. 1997, \aap, 317, L55

\bibitem[{{Dutrey} {et~al.}(2007){Dutrey}, {Guilloteau}, \&
  {Ho}}]{Dutrey+etal_2007}
{Dutrey}, A., {Guilloteau}, S., \& {Ho}, P. 2007, Protostars and Planets V, 495

\bibitem[{{Dutrey} {et~al.}(2011){Dutrey}, {Wakelam}, {Boehler}, {Guilloteau},
  {Hersant}, {Semenov}, {Chapillon}, {Henning}, {Pi{\'e}tu}, {Launhardt},
  {Gueth}, \& {Schreyer}}]{Dutrey+etal_2011}
{Dutrey}, A., {Wakelam}, V., {Boehler}, Y., {et~al.} 2011, \aap, 535, A104

\bibitem[{{Flock} {et~al.}(2011){Flock}, {Dzyurkevich}, {Klahr}, {Turner}, \&
  {Henning}}]{Flock+etal_2011}
{Flock}, M., {Dzyurkevich}, N., {Klahr}, H., {Turner}, N.~J., \& {Henning}, T.
  2011, \apj, 735, 122

\bibitem[{{Fromang} \& {Nelson}(2006)}]{Fromang+Nelson_2006}
{Fromang}, S. \& {Nelson}, R.~P. 2006, \aap, 457, 343

\bibitem[{{Gammie}(1996)}]{Gammie_1996}
{Gammie}, C.~F. 1996, \apj, 457, 355

\bibitem[{{Guilloteau} {et~al.}(2011){Guilloteau}, {Dutrey}, {Pi{\'e}tu}, \&
  {Boehler}}]{Guilloteau+etal_2011}
{Guilloteau}, S., {Dutrey}, A., {Pi{\'e}tu}, V., \& {Boehler}, Y. 2011, \aap,
  529, A105

\bibitem[{{Hartmann} {et~al.}(1998){Hartmann}, {Calvet}, {Gullbring}, \&
  {D'Alessio}}]{Hartmann+etal_1998}
{Hartmann}, L., {Calvet}, N., {Gullbring}, E., \& {D'Alessio}, P. 1998, \apj,
  495, 385

\bibitem[{{Henning}(2008)}]{Henning_2008}
{Henning}, T. 2008, Physica Scripta Volume T, 130, 014019

\bibitem[{{Henning} {et~al.}(2010){Henning}, {Semenov}, {Guilloteau}, {Dutrey},
  {Hersant}, {Wakelam}, {Chapillon}, {Launhardt}, {Pi{\'e}tu}, \&
  {Schreyer}}]{Henning+etal_2010}
{Henning}, T., {Semenov}, D., {Guilloteau}, S., {et~al.} 2010, \apj, 714, 1511

\bibitem[{{Hersant} {et~al.}(2009){Hersant}, {Wakelam}, {Dutrey}, {Guilloteau},
  \& {Herbst}}]{Hersant+etal_2009}
{Hersant}, F., {Wakelam}, V., {Dutrey}, A., {Guilloteau}, S., \& {Herbst}, E.
  2009, \aap, 493, L49

\bibitem[{{Horne} \& {Marsh}(1986)}]{Horne+Marsh_1986}
{Horne}, K. \& {Marsh}, T.~R. 1986, \mnras, 218, 761

\bibitem[{{Hueso} \& {Guillot}(2005)}]{Hueso+Guillot_2005}
{Hueso}, R. \& {Guillot}, T. 2005, \aap, 442, 703

\bibitem[{{Hughes} {et~al.}(2011){Hughes}, {Wilner}, {Andrews}, {Qi}, \&
  {Hogerheijde}}]{Hughes+etal_2011}
{Hughes}, A.~M., {Wilner}, D.~J., {Andrews}, S.~M., {Qi}, C., \& {Hogerheijde},
  M.~R. 2011, \apj, 727, 85

\bibitem[{{Ilgner} {et~al.}(2004){Ilgner}, {Henning}, {Markwick}, \&
  {Millar}}]{Ilgner+etal_2004}
{Ilgner}, M., {Henning}, T., {Markwick}, A.~J., \& {Millar}, T.~J. 2004, \aap,
  415, 643

\bibitem[{{Ilgner} \& {Nelson}(2006)}]{Ilgner+Nelson_2006}
{Ilgner}, M. \& {Nelson}, R.~P. 2006, \aap, 445, 223

\bibitem[{{Johansen} {et~al.}(2011){Johansen}, {Klahr}, \&
  {Henning}}]{Johansen+Klahr+Henning_2011}
{Johansen}, A., {Klahr}, H., \& {Henning}, T. 2011, \aap, 529, A62

\bibitem[{{Johansen} {et~al.}(2007){Johansen}, {Oishi}, {Mac Low}, {Klahr},
  {Henning}, \& {Youdin}}]{Johansen+etal_2007}
{Johansen}, A., {Oishi}, J.~S., {Mac Low}, M.-M., {et~al.} 2007, \nat, 448,
  1022

\bibitem[{{Keller} \& {Gail}(2004)}]{Keller+Gail_2004}
{Keller}, C. \& {Gail}, H.-P. 2004, \aap, 415, 1177

\bibitem[{{Lynden-Bell} \& {Pringle}(1974)}]{LindenBell+Pringle_1974}
{Lynden-Bell}, D. \& {Pringle}, J.~E. 1974, \mnras, 168, 603

\bibitem[{{Oishi} {et~al.}(2007){Oishi}, {Mac Low}, \&
  {Menou}}]{Oishi+etal_2007}
{Oishi}, J.~S., {Mac Low}, M.-M., \& {Menou}, K. 2007, \apj, 670, 805

\bibitem[{{Papaloizou} \& {Terquem}(2006)}]{Papaloizou+Terquem_2006}
{Papaloizou}, J.~C.~B. \& {Terquem}, C. 2006, Reports on Progress in Physics,
  69, 119

\bibitem[{{Pi{\'e}tu} {et~al.}(2007){Pi{\'e}tu}, {Dutrey}, \&
  {Guilloteau}}]{Pietu+etal_2007}
{Pi{\'e}tu}, V., {Dutrey}, A., \& {Guilloteau}, S. 2007, \aap, 467, 163

\bibitem[{{Prandtl}(1925)}]{Prandtl_1925}
{Prandtl}, L. 1925, Zs. angew. Math. Mech., 5, 136

\bibitem[{{Regev} \& {Gitelman}(2002)}]{Regev+Gitelman_2002}
{Regev}, O. \& {Gitelman}, L. 2002, \aap, 396, 623

\bibitem[{{Semenov} {et~al.}(2008){Semenov}, {Pavlyuchenkov}, {Henning},
  {Wolf}, \& {Launhardt}}]{Semenov+etal_2008}
{Semenov}, D., {Pavlyuchenkov}, Y., {Henning}, T., {Wolf}, S., \& {Launhardt},
  R. 2008, \apjl, 673, L195

\bibitem[{{Semenov} \& {Wiebe}(2011)}]{Semenov+Wiebe_2011}
{Semenov}, D. \& {Wiebe}, D. 2011, \apjs, 196, 25

\bibitem[{{Semenov} {et~al.}(2006){Semenov}, {Wiebe}, \&
  {Henning}}]{Semenov+etal_2006}
{Semenov}, D., {Wiebe}, D., \& {Henning}, T. 2006, \apjl, 647, L57

\bibitem[{{Simon} {et~al.}(2011){Simon}, {Armitage}, \&
  {Beckwith}}]{Simon+Armitage+Beckwith_2011}
{Simon}, J.~B., {Armitage}, P.~J., \& {Beckwith}, K. 2011, \apj, 743, 17

\bibitem[{{Tscharnuter} \& {Gail}(2007)}]{Tscharnuter+Gail_2007}
{Tscharnuter}, W.~M. \& {Gail}, H.-P. 2007, \aap, 463, 369

\bibitem[{{Uribe} {et~al.}(2011){Uribe}, {Klahr}, {Flock}, \&
  {Henning}}]{Uribe+etal_2011}
{Uribe}, A.~L., {Klahr}, H., {Flock}, M., \& {Henning}, T. 2011, \apj, 736, 85

\bibitem[{{Urpin}(1984)}]{Urpin_1984}
{Urpin}, V.~A. 1984, \sovast, 28, 50

\bibitem[{{Voelk} {et~al.}(1980){Voelk}, {Jones}, {Morfill}, \&
  {Roeser}}]{Voelk+etal_1980}
{Voelk}, H.~J., {Jones}, F.~C., {Morfill}, G.~E., \& {Roeser}, S. 1980, \aap,
  85, 316

\end{thebibliography}
\bibliographystyle{aa}

\end{document}